\documentclass[aip,rsi,reprint]{revtex4-1}
\usepackage{graphicx}
\usepackage{amsmath}
\usepackage{graphicx}
\usepackage[squaren]{SIunits}

\usepackage{dcolumn}
\usepackage{bm}

\begin{document}

\title{A scanning probe-based pick-and-place procedure for assembly of integrated quantum optical hybrid devices}

\author{Andreas W. Schell}
\email[Electronic mail: ]{andreas.schell@physik.hu-berlin.de}
\affiliation{Nano-Optics, Institute of Physics, Humboldt-Universit\"{a}t zu
Berlin, Newtonstra{\ss}e~15, D-12489 Berlin, Germany}

\author{G\"{u}nter Kewes}
\affiliation{Nano-Optics, Institute of Physics, Humboldt-Universit\"{a}t zu
Berlin, Newtonstra{\ss}e~15, D-12489 Berlin, Germany}

\author{Tim Schr\"{o}der}
\affiliation{Nano-Optics, Institute of Physics, Humboldt-Universit\"{a}t zu
Berlin, Newtonstra{\ss}e~15, D-12489 Berlin, Germany}

\author{Janik Wolters}
\affiliation{Nano-Optics, Institute of Physics, Humboldt-Universit\"{a}t zu
Berlin, Newtonstra{\ss}e~15, D-12489 Berlin, Germany}

\author{Thomas Aichele}
\affiliation{Nano-Optics, Institute of Physics, Humboldt-Universit\"{a}t zu
Berlin, Newtonstra{\ss}e~15, D-12489 Berlin, Germany}

\author{Oliver Benson}
\affiliation{Nano-Optics, Institute of Physics, Humboldt-Universit\"{a}t zu
Berlin, Newtonstra{\ss}e~15, D-12489 Berlin, Germany}

\begin{abstract}
Integrated quantum optical hybrid devices consist of fundamental constituents such as single emitters and
tailored photonic nanostructures. A reliable fabrication method requires the controlled deposition of active
nanoparticles on arbitrary nanostructures with highest precision. Here, we describe an easily adaptable
technique that employs picking and placing of nanoparticles with an atomic force microscope combined with a
confocal setup. In this way, both the topography and the optical response can be monitored simultaneously before
and after the assembly. The technique can be applied to arbitrary particles. Here, we focus on nanodiamonds
containing single nitrogen vacancy centers, which are particularly interesting for quantum optical experiments
on the single photon and single emitter level.
\end{abstract}

\maketitle

\section{Introduction}

In recent years, the integration of single quantum emitters into nanophotonic structures such as microcavities
\cite{Vahala2003}, optical antennas \cite{Muhlschlegel2005} or waveguides \cite{OBrien2009} has attracted major
interest in quantum and nano optics. Especially nitrogen vacancy (NV) defect centers in diamond crystals turned
out to be stable and bright single-photon emitters even under ambient conditions
\cite{Kurtsiefer2000,Brouri2000}. Due to the triplet ground state with electron spin decoherence times in the
millisecond range, NV centers are also used as quantum memory systems and as nanomagnetic probes
\cite{2009NatMa...8..383B,2008Natur.455..648B}. Color centers occur naturally or can be artificially inserted
into the diamond crystal by ion implantation \cite{meijer:261909,Meijer2008}. Combined with lithographic
techniques, this top-down approach allows permanent integration of color centers into microcavities and
waveguides \cite{babinec:010601,Fu2008}. In contrast, diamond nanocrystals containing single defect centers can
be coupled to photonic nanostructures to build hybrid quantum systems in a bottom-up approach. For positioning
the nanocrystals with nanometer precision a scanning electron microscope (SEM) with a manipulator
\cite{Ampem-Lassen2009,Sar2009} or a scanning atomic force microscope (AFM)
\cite{Schietinger2009,Barth2009,PhysRevLett.106.096801} have been used to date.

Controlled pushing of nanometer-sized objects on a surface through the AFM tip has been demonstrated for the
first time by Junno et al. \cite{Junno1995} Picking up of metallic nanoparticles can be achieved by using
electrostatic forces \cite{Toset2007} or after chemical treatment of the AFM tip \cite{Tanaka2010}. In previous
experiments \cite{Wolters2010,Schroder2010,Schell2011} we were striving for a refined dip-pen technique
\cite{Wang2007} for assembly of fluorescent nanoparticles on a multitude of optical devices. In this paper, we
now describe an easily adaptable pick-and-place procedure which is particularly suitable for precise positioning
of diamond nanocrystals.

\begin{figure*}
  \includegraphics{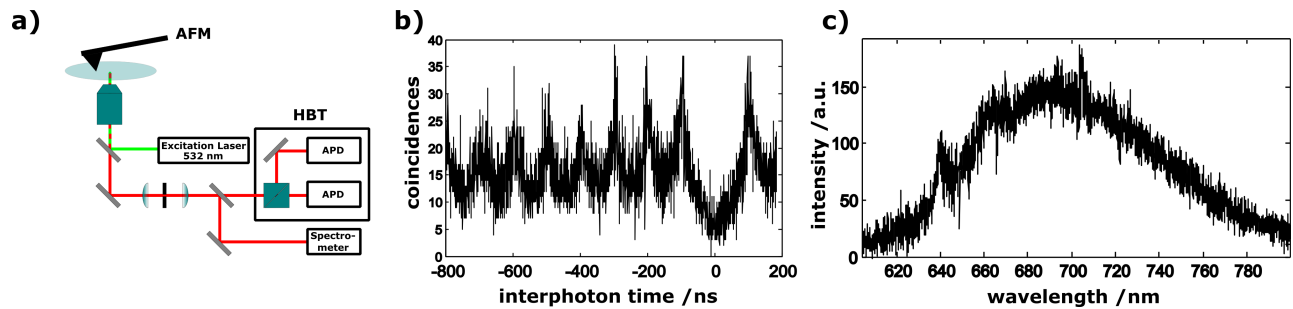}
  \caption{(a) Schematics of the setup used to pre-characterize and pick up nanodiamonds. (b) Example of an autocorrelation
  measurement of fluorescence from an NV center in a nanodiamond showing pronounced antibunching. The repetition rate of the pulsed
  excitation laser at 532 nm was $\unit{10}{\mega\hertz}$. (c) Spectrum of the same NV center with a zero phonon line peak at
  $\unit{639.6}{\nano\meter}$.}
  \label{fig:setup}
\end{figure*}

\begin{figure}
  \includegraphics{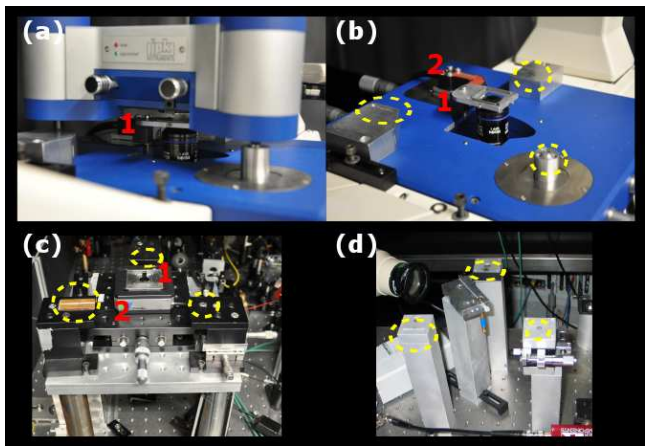}
  \caption{Photograph the body of a Zeiss Axiovert 200 microscope converted to a confocal microscope with the AFM atop (a)
  and removed (b), respectively.
  (c) Top-view photograph of the homebuilt confocal microscope. (d) Detailed view of a special holder consisting of
  three posts for mounting the AFM to approach complex shaped items, e.g., an optical fiber.
  The dashed circles label the AFM mounts, 1 labels the sample holder and
  2 labels the piezo sample scanner.}
  \label{fig:aufbauten}
\end{figure}

\section{Experimental Setup}

The experimental setup (Figure \ref{fig:setup} (a)) for our pick-and-place process consists of an inverse
confocal microscope with an AFM (NanoWizard, JPK Instruments) atop. We used two different microscopes. One is 
a Zeiss Axiovert 200 (Figure \ref{fig:aufbauten} (a) and (b)), the other one was homebuilt (Figure
\ref{fig:aufbauten} (c)). Additionally, a special holder was constructed to allow for AFM manipulation on more
complex or fragile photonic structures, such as optical fibers (Figure \ref{fig:aufbauten} (d)). While the AFM
is a tip scanner with three axes, the confocal microscope has a 2D piezo sample stage (either PXY 80 D12,
piezosystem jena / PXY100 ID, piezosystem jena) and a piezo actuated z-axis objective positioning system (MIPOS
100, piezosystem jena). In this way, the nanoparticle sample as well as the AFM tip can be positioned
independently relative to the laser focus. 

A sample was produced by spincoating of an ensemble of nanodiamonds
from a solution on a glass coverslip. The solution is a suspension of centrifuge cleaned nanodiamonds
(Microdiamant AG) in water with 0.02\% polyvinyl alcohol. On such a sample individual nanodiamonds were
pre-characterized prior to the pick up procedure. Light from a pulsed laser (LDH-P-FA-530, PicoQuant) with a
wavelength of $532\,\rm nm$ and a repetition rate of $10\,\rm MHz$ was focussed on a nanodiamond via a high
numerical aperture objective (UPlanSApo 60XO, Olympus / PlanApo 60XO, Olympus). Its fluorescence was dispersed
by a grating spectrometer (SpectraPro-2500i, Acton) to identify a characteristic NV spectrum (see Figure
\ref{fig:setup} (c)). Autocorrelation measurements of the fluorescence were performed with the help of a Hanbury
Brown and Twiss (HBT) setup consisting of a 50/50 beam splitter and two avalanche photodiodes (PDM Series, Micro
Photon Devices / SPCM-AQR-14, Perkin Elmer). Time intervals between photons were measured with either a TimeHarp
200 or a Picoharp 300 from PicoQuant. By evaluating the autocorrelation function $g^{(2)}(\tau)$ (see Figure
\ref{fig:setup} (b)) at $\tau=0$ the number of emitting NV centers in the nanodiamond could be determined
\cite{Sonnefraud2008}. Only nanodiamonds containing a single NV center, i.e., those with a vanishing peak at
$g^{(2)}(0)=0$ were used for the subsequent pick-and-place procedure. 

A homebuilt nanosecond pulse counter was
used for monitoring the optical signal and for converting the digital signal of the APDs to an analog voltage
which was fed to the analog to digital converter of the AFM. This provides the opportunity to directly overlay 
topography and optical signal. The AFM was controlled with its standard software
while homemade software written in Labview (National Instruments) and a multi-function data acquisition card
(PCI-6014, National Instruments) were used to control the confocal microcope.

\section{The Pick-and-Place Procedure}

The pre-characterized nanodiamond is placed into the optical focus of the confocal microscope and is identified
with the AFM by scanning the tip over the focus in intermittent contact mode. In addition to the standard AFM
images like topography and phase we also record the optical signal from the optical microscope versus tip
position. To suppress the excitation light we used a longpass filter at $\lambda=\unit{590}{\nano\meter}$. With
the AFM approached we used an additional shortpass filter at $\lambda=\unit{740}{\nano\meter}$ to suppress the infrared AFM laser. 

\begin{figure*}
  \includegraphics{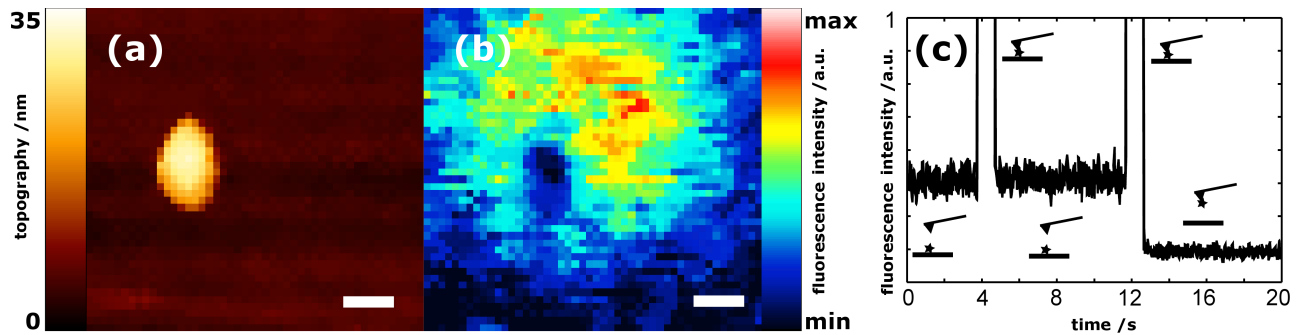}
  \caption{(a) AFM topography image of a nanodiamond in the spot of a confocal microscope's laser.
  (b) Optical image, i.e., detected fluorescence signal versus tip position. In this measurement the collected
  fluorescence is reduced when the tip scans across the diamond nanoparticle (see text). Scalebars in (a,b) are $\unit{100}{\nano\meter}$.
  (c) Detected fluorescence signal when picking up the diamond. The fluorescence increases when the tip is at the
  sample surface. After a first unsuccessful attempt where the fluorescence had fully recovered the pick up procedure was
  repeated, and finally the nanodiamond was picked up indicated by a drop of the fluorescence signal to the background level.}
  \label{fig:afmoptical}
\end{figure*}

The optical
signal consists of two contributions. Firstly, there is a constant fluorescence signal from the NV center in the
laser focus. A second contribution stems from fluorescence of the AFM tip which depends on the position of tip
relative to the focus. Thus, scanning the tip over the laser focus results in an AFM topography image together
with an optical image of the focus area. Figure~\ref{fig:afmoptical} (a,b) show the AFM topography and the
optical image, respectively, with a single nanodiamond in the laser focus. In some cases (Figure
\ref{fig:afmoptical} (b)) the fluorescence drops at the nanodiamond location. This is due to a modified
scattering of the tip's fluorescence towards the collection optics of the confocal microscope when the tip is
scanned across the nanodiamond. If the density of nanodiamonds on the substrate is sufficiently low, a single
diamond nanoparticle can be identified in the laser focus unambiguously.

The pick up procedure is started by positioning the AFM tip above the nanodiamond. Then, the tip is pressed on
the center of the particle in contact mode. A force of up to \unit{1}{\micro\newton} is applied which is
sufficient to attach the particle to the tip due to surface adhesion. Simultaneously the fluorescence is
observed. If the nanodiamond is picked up successfully, the fluorescence signal drops to background level after
the tip is retracted (see Figure \ref{fig:afmoptical} (c)). In order to ensure that the nanodiamond is picked up
by the tip, and not only pushed out of the laser focus, the sample stage is used to scan the vicinity of the
original nanodiamond position. If the pick up was not successful the tip is pressed on the nanodiamond again
until it is finally picked up. From time to time, an additional topography image with the AFM in intermittent
contact mode is taken in order to determine the diamond's position. This is necessary, because the diamond
sometimes moves a distance on the order of the tip radius when touched by the AFM tip. In our experiments a pick
up was always possible, even if it could take a large number of approaches (sometimes over 50).

After being picked up the nanodiamond can be transferred to any structure accessible with the AFM. It is even
possible to transport the whole AFM to another setup without losing the nanodiamond. If the new structure is not
suitable for confocal microscopy with simultaneous AFM access, care has to be taken that the diamond can be
clearly identified after it has been deposited. Therefore, a small area (e.g. $\unit{0.1}{\micro\meter}^2$) on
the targeted structure is scanned by the AFM in intermittent contact mode. In this scanning process it is
unlikely to lose the diamond as long as there are no sharp edges on the target surface. The diamond is then
deposited by pressing the tip on the surface with a force of up to \unit{1}{\micro\newton} and the area is
scanned again. This is repeated until the nanodiamond appears on the topography image. 

In contrast to the pick
up process this is not always successful. Only approximately one third of the diamonds picked up could be placed again.
We attribute this to nanodiamonds sticking at the side of the tip instead of the tip apex. When pressed to the surface, these
nanodiamonds are pushed further along the side of the tip until they can not reach the surface anymore.
Obviously, there is always a competition among adhesion between the nanoparticle and the tip and the
nanoparticle and the target surface, respectively. When a dimanond was lost, a new cantilever was used to make sure
that the diamond deposited is really the one pre-characterized before.

\begin{figure}
  \includegraphics{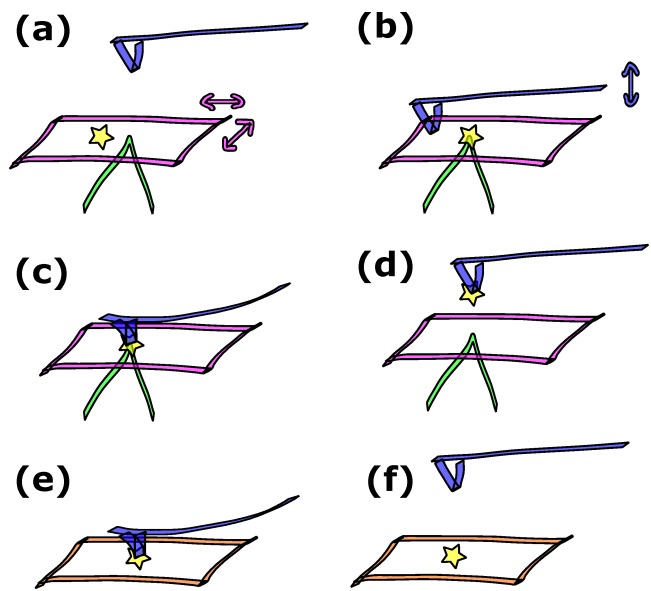}
  \caption{Scheme of the nanodiamond pick-and-place procedure. (a) The sample is scanned in the confocal microscope in
  order to find and optically characterize a nanodiamond. (b) The AFM tip is scanned across the focal region of the microscope to
identify the chosen nanodiamond. (c) The tip is pressed on the nanodiamond. (d) The nanodiamond sticks to the
tip.
  (e) The tip is pressed on a new structure to deposit the nanodiamond. (f) The diamond is positioned at the desired
  position.}
  \label{fig:pickscheme}
\end{figure}

A sketch of the whole procedure is given in Figure \ref{fig:pickscheme}. The technique is presented here for
nanodiamonds. In principle, it is possible to extend it to any other nanoparticle since it only relies on
surface adhesion and does not require a chemical functionalization of the surfaces.

\begin{figure}
  \includegraphics{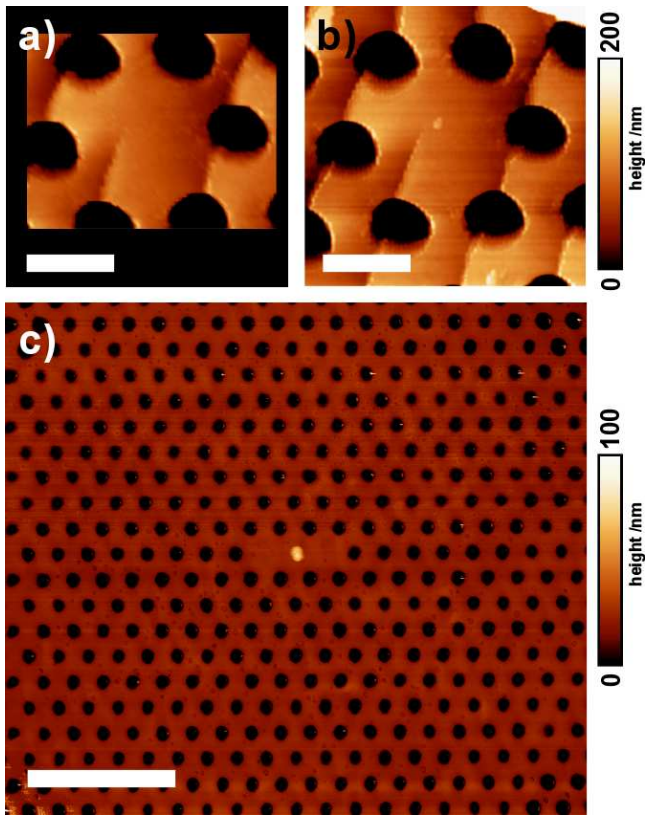}
  \caption{(a) and (b) AFM image of the core of a photonic crystal fiber before and after placing a nanodiamond, respectively.
  (c) A nanodiamond placed inside a gallium phosphide photonic crystal membrane cavity. The thickness
  of the free-standing membrane is approx. \unit{60}{\nano\meter}. All scalebars are $\unit{1}{\micro\meter}$.}
  \label{fig:fibreandcavity}
\end{figure}

The pick-and-place procedure has to be refined if the targeted structures have sharp edges near the desired
nanoparticle position. Examples are photonic crystal cavities \cite{Wolters2010} or photonic crystal fibers
\cite{Schroder2010}. In this case, a two-step process is needed. The nanodiamond is first placed on a smoother
area of the target structure. Then, an AFM topography image of the targeted region can be taken with the bare
tip. In this way the risk of losing the nanodiamond when scanning tip and nanodiamond across sharp edges is
avoided. With the targeted region well identified via the AFM topography image, the diamond is finally
transferred to its target position by a second pick-and-place process. One disadvantage of this two-step process
is the lack of optical control during the second pick up, what makes the whole process more time consuming,
since after each try an AFM scan has to be performed in order to determine if the nanodiamond has been picked
up. Two examples of nanodiamonds transfered with the two step process are shown in Figure \ref{fig:fibreandcavity}
(a-c).

\begin{figure}
  \includegraphics{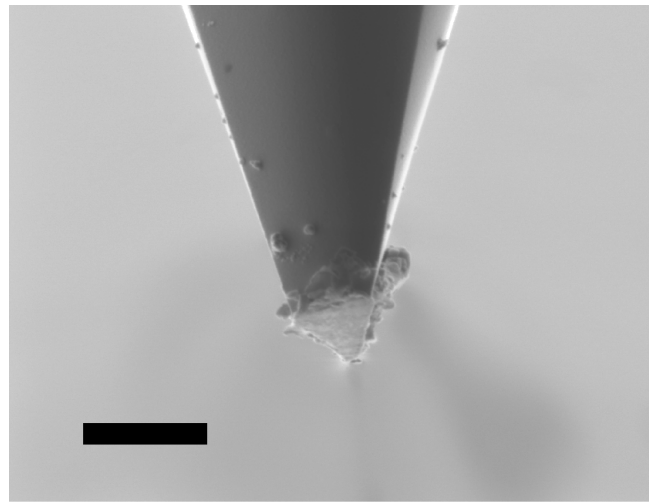}
  \caption{SEM image of Pt/Ti coated cantilever used for the pick-and-place procedure. The tip has flattened
  by being pressed on the surface in order to pick up a nanodiamond.
  Scalebar is $\unit{1}{\micro\meter}$.}
  \label{fig:cantilever}
\end{figure}

In principle, the pick-and-place procedure can be performed with any AFM cantilever, but for optimum
performance, there are some requirements. First, it is advantageous for the cantilever tip to have a radius of
curvature which is large, since the probability for the nanodiamond to attach to the tip's side rather than to
its apex decreases with increasing radius. On the other hand, the radius of curvature has to be sufficiently
small to identify single nanoparticles in an AFM topography image. Second, ductile tips are preferred because
they do not break when being pressed multiple times on the sample. Third, the tip material is important, because
the adhesion forces strongly depend on the involved materials \cite{Eastman1996}. To our experience these requirements are best
met by metal coated silicon tips, which are commercially available (we used Au and Pt/Ti coated cantilevers from
MicroMasch). These tips seldom break compared to uncoated ones, have a higher radius of curvature (approx.
\unit{40}{\nano\meter}) and it is possible to deform them by pressing them on the substrate or on a nanodiamond.
An example for a used Pt/Ti coated tip can be seen in Figure \ref{fig:cantilever}.

\section{Conclusion}

In conclusion, we have described a versatile technique to transfer nanoparticles, in particular nanodiamonds,
with nanometer precision even between different samples. This technique allows positioning of specific
nanoparticles to a variety of structures, overcoming high densities and random positioning during a spin-coating
operation. The method is particularly attractive for samples where a standard spin-coating of nanoparticles does
not work for geometrical reasons, like optical fibers \cite{Schroder2010} and nanocavities \cite{Barth2009}.
Moreover the presented technique can be used at ambient environments and can be combined with optical monitoring
in contrast to manipulation in a SEM.

\section*{Acknowledgement}

We acknowledge financial support from DFG through project IQuOSuPla (AI 92/3) and thank Max Schoengen for
taking the SEM picture in Figure \ref{fig:cantilever}.

\end{document}